\documentclass[10pt]{article}
\usepackage{amsmath,amsthm,amssymb,amsfonts, fancyhdr, color, comment, graphicx, environ, esint, float, mdframed, stix, subcaption, indentfirst}
\usepackage[breakable]{tcolorbox}
\usepackage[top=1in, bottom=1in, left=1in, right=1in]{geometry}
\newcommand{\supp}{\mathrm{supp}}

\title{On Non-uniqueness of Phase Retrieval in Multidimensions}
\author{by Roman G. Novikov and Tianli Xu}
\date{}
\begin{document}
\maketitle
\noindent \textbf{Abstract}: We give a large class of examples of non-uniqueness for the phase retrieval problem in multidimensions. Our constructions are based on "oblique tensorization", where one-dimensional results are strongly used, and its generalizations towards complete description of non-uniqueness. Our examples include the case of functions with strongly disconnected compact support. 
\\ \hfill 
\section{Introduction} 
The phase retrieval problem consists in finding a function $v: \mathbb{R}^d \rightarrow \mathbb{C}$ from the magnitude $|\hat{v}|$ of its Fourier transform 
\begin{equation}
    \hat{v}(p) = Fv(p) = \frac{1}{(2 \pi)^d}\int_{\mathbb{R}^d} e^{ipx}v(x) dx \text{, }  p \in \mathbb{R}^d.
\end{equation}

\indent This problem naturally arises in quantum mechanics, optics, and related areas such
as electron tomography and X-ray imaging; see, for example, [1] - [4], [6] - [9], [11] - [13], [15] and references therein. \\
\indent In general, many different functions $v$ have the same Fourier modulus.
These different solutions can be obtained by multiplying $|\hat{v}|$ by measurable complex-valued functions with modulus one and taking the inverse Fourier transform; see, for example, [3]. \\
\indent When $v$ is compactly supported, the degree of ambiguity is reduced. In particular, for $d=1$,  all solutions with compact support could be obtained from any
one of them by flipping (conjugating) non-real zeros of its Fourier transform extended by analyticity to the complex plane; see [15], [4]. \\
\indent When $v$ is a sum of functions with sufficiently disconnected compact supports, the degree of ambiguity is further reduced. In particular, for $d = 1$, this ambiguity is completely described in [3]. Roughly speaking, in this case, the phase retrieval problem almost always has essentially a unique solution. Nevertheless, important examples of non-uniqueness for this case are also given in [2], [7]. Note that the examples of [7] admit analogs in multidimensions. \\
\indent In addition, it is also mentioned in the literature that for functions with compact support, the degree of non-uniqueness of phase retrieval is further reduced in dimension $d \geq 2$, in general, and for the case of sufficiently disconnected support in particular; see [3]. This further reduction of non-uniqueness in dimension $d \geq 2$ can be also related to the result that multi-variable polynomials are known to be generically irreducible; see [13], chapter 4 of [1] and, for example, section 2 of [7] for related results and discussions in the framework of phase retrieval.  \\
\indent Moreover, the important work [9] suggests an efficient numerical phase retrieval algorithm for functions with sufficiently disconnected compact support. This algorithm works very well numerically (at least, for $d = 2$) and possible non-trivial non-uniqueness is not even discussed in [9]. \\
\indent Recall that the non-uniqueness in phase retrieval for compactly supported $v$ with possible additional assumptions is non-trivial (and of interest) if it does not reduce to the functions $v_{\alpha, y}$ and $\tilde{v}_{\alpha, y}$ associated to $v$, where
\begin{equation}
   v_{\alpha, y}(x) = e^{i \alpha} v(x - y),
\end{equation}
\begin{equation}
   \tilde{v}_{\alpha, y}(x) = e^{i \alpha}\overline{v(-x + y)},
\end{equation}
and $\alpha \in \mathbb{R}$, $y \in \mathbb{R}^d$, bar denotes the complex conjugation; see, for example, [2]. \\
\indent Recall also that via tensorization, i.e., in the framework of functions of the form 
\begin{equation}
    v(x) = \prod_{j = 1}^dv_j(x_j), \, x = (x_1, \cdots, x_d) \in \mathbb{R}^d,
\end{equation}
it is extremely easy to build examples for which the solution of the multidimensional phase retrieval problem is not unique and does not reduce to associated functions, using the aforementioned one-dimensional results of [15], [4]. \\ 
\indent The results of the present article can be summarized as follows. \\\indent First, we construct a very large class of non-trivial examples of non-uniqueness for phase retrieval in multidimensions via "oblique tensorization" and results of [15], [4]; see Theorem 3.1 in Section 3. \\
\indent Second, we strongly generalize the oblique tensorization construction in Theorems 3.2 and further in Theorem 3.3 in Section 3, where we do not use necessarily one-dimensional results of [15], [4]. In this connection, our considerations also involve ideas going back to [13].\\
\indent Third, proceeding from Theorems 3.1, 3.2, and 3.3, we give non-trivial examples of non-uniqueness for phase retrieval for functions with strongly disconnected compact support in muldidimensions; see Theorems 3.4 and 3.5 in Section 3. \\
\indent Thus, there are many important examples of non-uniqueness for phase retrieval in multidimensions, even for functions with strongly disconnected support. In this connection, the conventional opinion that such non-uniqueness is not essential and is negligible seems to be overstated. Thus, in order to have complete uniqueness in phase retrieval for functions $v$ even with strongly disconnected compact support, and even modulo associated functions in dimension $d \geq 2$, additional a priori information is necessary. In connection with natural theoretical and numerical results in this direction, see, for example, [1], [6], [11] and references therein. \\
\indent A preliminary version of this article corresponds to the preprint [12]. 
 \section{Preliminaries}
 \indent Note that formulas (2), (3) can be rewritten as follows in the Fourier domain: 
\begin{equation}
    \hat{v}_{\alpha, y}(p) = \hat{v}(p)\operatorname{exp}(i\alpha + iyp) \text{, } p \in \mathbb{R}^d,
\end{equation}
\begin{equation}
    \hat{\tilde{v}}_{\alpha, y}(p) = \overline{\hat{v}(p)}\operatorname{exp}(i\alpha + iyp) \text{, } p \in \mathbb{R}^d.
\end{equation}

 \indent Recall that 
\begin{equation}
    F^{-1}(\phi_1 \phi_2) = (2 \pi)^{-d} F^{-1}\phi_1 * F^{-1}\phi_2,
\end{equation}
\begin{equation}
    (2\pi)^d F(u_1 * u_2) = Fu_1 Fu_2,
\end{equation}
 where $F$ is defined by (1), $\phi_1$, $\phi_2$, $u_1$, $u_2$ are test functions on $\mathbb{R}^d$, and $*$ denotes the convolution, that is, 
\begin{equation}
    u_1 * u_2(x) = \int_{\mathbb{R}^d} u_1(x -y) u_2(y) dy \text{, } x \in \mathbb{R}^d.
\end{equation}
 \indent Let 
 \begin{equation}
    B_r = \{ x \in \mathbb{R}^d: |x| \leq r\} \text{, }r > 0,
\end{equation}
\begin{equation}
    H(x) = \mathrm{sup}_{\substack{y \in K} } \, xy, \, x \in \mathbb{R}^d,
\end{equation}
 where $K$ is a convex compact subset of $\mathbb{R}^d$. Note that $H(x) = r|x|$ if $K = B_r$.
\\ \hfill \\
 \indent In the framework of phase retrieval analysis, the following theorem is essential. 
 \\ \hfill \\
 \indent \textbf{Theorem 2.1 (Paley-Winer-Schwartz)}: \textit{Let $K$ be a convex compact subset of $\mathbb{R}^d$. An entire function $\hat{v}$ on $\mathbb{C}^d$ is the Fourier transform of a distribution $v$ supported in $K$ if and only if 
 \begin{equation}
     |\hat{v}(p)| \leq C(1 + |p|)^m e^{H(\mathrm{Im}(p))}, \, p \in \mathbb{C}^d,
\end{equation}
where $C \geq 0$ and $m \in \mathbb{R}$ are some constants, $H$ is defined by (11). }
\\ \hfill \\
\indent Recall also that $v \in L^2(\mathbb{R}^d)$ and $\supp \, v \subseteq K$ if and only if estimate (12) holds for $m = 0$ and $\hat{v} \in L^2(\mathbb{R}^d)$, where $K$ is as in Theorem 2.1; this result is known as Paley-Winer-Plancherel-Polya theorem.\\
\indent In connection with Theorem 2.1, see, for example, Theorem 7.3.1 in [5]. 
\\ \hfill \\
\indent At least for $d = 1$, the phase retrieval analysis strongly uses the following Hadamard's factorization formula:
\begin{equation}
    \hat{f}(z) = e^{c_0 + c_1z} z^k \prod_{\zeta \in Z_f} \left(1 - \frac{z}{\zeta}\right)e^{z/\zeta},\, z \in \mathbb{C},
\end{equation}
where $f \in L^2(\mathbb{R)}$ is a compactly supported function, $\hat{f} \nequiv 0$, and $Z_f$ denotes the set of the zeros of $\hat{f}$, where each zero is repeated according to its multiplicity, and $c_0$, $c_1 \in \mathbb{C}$, $k \in \mathbb{N} = \{0, 1, 2, \cdots\}$; see, for example, Theorem 8.2.4 in~[14].
\\ \hfill \\
\indent A fundamental result on phase retrieval for $d = 1$ consists of the following theorem. 
\\ \hfill \\
\indent \textbf{Theorem 2.2 (Walther)}: \textit{Let $f$ be as in (13). Then $|\hat{g}|^2 = |\hat{f}|^2$, where $g \in L^2(\mathbb{R})$ is a compactly supported function, if and only if 
\begin{equation}
    \hat{g}(z) = e^{ia + ibz}e^{c_0 +c_1 z} z^k \prod_{\zeta \in Z_g} \left(1 - \frac{z}{\zeta}\right)e^{z/\zeta},\, z \in \mathbb{C},
\end{equation}
where $a$, $b \in \mathbb{R}$, $i = \sqrt{-1}$ is the complex unit, and
\begin{equation}
    Z_g = \left(Z_f \setminus Z\right) \bigcup \overline{Z},
\end{equation}
 $Z$ is a subset of non-real zeros of $Z_f$. In addition, presentation (14) can be rewritten as 
\begin{equation}
    \hat{g}(z) = \operatorname{exp}(i\alpha + i\beta z)\prod_{\zeta \in {Z}} \frac{1 - z/\bar{\zeta}}{1- z/ {\zeta}}\hat{f}(z), \text{ } \text{ } z \in \mathbb{C}, \text{ at least if } \# Z < + \infty,
\end{equation}
 where $\alpha$, $\beta \in \mathbb{R}$, and $\#$ denotes the cardinality of a set. Moreover, if $\supp \, f \subseteq [- \epsilon, \epsilon]$, and $\hat{g}$ is given by (16) where $\beta = 0$, 
then $\supp \, g \subseteq [- \epsilon, \epsilon]$.}
\\ \hfill \\
\indent Note that in Theorem 2.2, we have that $Z_g$ coincides with the set of zeros of $\hat{g}$. In addition, the case ${Z}_g = {Z}_f$ coincides with the case when $f$, $g $ are associated functions in the sense of formulas (2), (5), whereas the case ${Z}_g = \overline{{Z}_f}$ coincides with the case when $f$, $g$ are associated functions in the sense of formulas (3), (6).\\
\indent In connection with Theorem 2.2, see [15], [4], [7]. \\
\indent Note that in dimension $d \geq 2$, a proper analog of Theorem 2.2 is not yet available. In this case, key difficulties are related to the facts that the zero sets of Fourier transforms of compactly supported functions are no longer discrete and that multi-variable polynomials $P(z)$ are known to be generically irreducible;  see, for example, [13], chapter 4 of [1] and section 2 of [7]. Nevertheless, roughly speaking, it is known that if one has two compactly supported functions or even distributions $f$, $g$ on $\mathbb{R}^d$, then $|\hat{f}|^2 = |\hat{g}|^2$ on $\mathbb{R}^d$ if and only if  
\begin{equation}
    \hat{f}(z) = P(z)Q(z),\, z \in \mathbb{C}^d,
\end{equation}
\begin{equation}
    \hat{g}(z) = \operatorname{exp}(i\alpha + i\beta z)\overline{P\left(\overline{z}\right)} Q(z), \, z \in \mathbb{C}^d,
\end{equation}
where $P$, $Q$ are some holomorphic functions of exponential type on $\mathbb{C}^d$, $\alpha \in \mathbb{R}$ and $\beta \in \mathbb{R}^d$; see, for example, [13], [7] and references therein. Via Theorems 3.3 and 3.5 in Section 3, this observation leads to a method for constructing a large class of non-trivial examples of non-uniqueness for phase retrieval in multidimensions, in general, and for the case of strongly disjoint support, in particular. 
\\ \hfill \\
\indent Recall also that if $u_1$, $u_2$ are compactly supported distributions on $\mathbb{R}^d$, then 
\begin{equation}
    \mathrm{ch} \, \supp \text{ } u_1 * u_2 = \mathrm{ch} \, \supp \, u_1 + \mathrm{ch} \, \supp \, u_2,
\end{equation}
 where $\mathrm{ch} \, A$ is the convex hull of a set $A$ in $\mathbb{R}^d$ and 
\begin{equation}
    {A} + {B} = \{ x + y: \text{ } x \in {A}, \text{ } y \in {B}\} 
\end{equation}
\indent if $A$ and $B$ are subsets of $\mathbb{R}^d$; see [10].\\
\indent In addition,
\begin{equation}
    B_{r_1}(a_1) + B_{r_2}(a_2) = B_{r_1 + r_2}(a_1 +a_2), 
\end{equation}
 where 
\begin{equation}
    B_r(a) = \{ x \in \mathbb{R}^d: \text{ } |x - a| \leq r\}, \text{ } a \in \mathbb{R}^d, \text{ } r > 0. 
\end{equation}

\section{The main results}
\subsection{Non-uniqueness of phase retrieval}
First, we construct a large class of non-trivial examples of non-uniqueness for phase retrieval in dimension $d \geq 2$ via "oblique tensorization".\\
\indent Let 
\begin{equation}
    f_{j}, g_{j} \in L^2(\mathbb{R}), \, \supp \, f_{j}, \, \supp \, g_{j} \subseteq [-\epsilon, \epsilon], \, \epsilon>0, \, |\hat{f}_{j}|^2 = |\hat{g}_{j}|^2 \nequiv 0 \text{ on } \mathbb{R}^d,\, j = 1, \cdots, N,
\end{equation}
 where $f_{j}$ and $g_{j}$ are as $f$, $g$ in Theorem 2.2 with ${Z}_{f_j}$, ${Z}_{g_j}$, ${Z}_j$ in place of ${Z}_f$, ${Z}_g$, ${Z}$. In addition, we assume that
\begin{equation}
  {Z}_{g_j} \neq  {Z}_{f_j}, \, {Z}_{g_j} \neq \overline{{Z}_{f_j}}.
\end{equation} 
\indent For $d \geq 2$, we define $f$, $g$ as
\begin{equation}
       f = F^{-1}\hat{f},\indent  \hat{f}(p) = \prod_{j = 1} ^N \hat{f}_j(\omega_j \cdot p), \text{ } p \in \mathbb{R}^d, 
\end{equation}
\begin{equation}
       g = F^{-1}\hat{g},\indent  \hat{g}(p) = \prod_{j = 1} ^N \hat{g}_j(\omega_j \cdot p),\text{ } p \in \mathbb{R}^d, 
\end{equation}
where $\omega_j \in \mathbb{S}^{d-1}$, and $\omega_j \neq \pm \omega_k$ if $j \neq k$. 
\\ \hfill \\
\indent Let $B_r$ be defined by (10).
\\ \hfill \\
\indent \textbf{Theorem 3.1}: \textit{Let $f$, $g$ be defined by (25), (26). Then: \\
\indent (i) $\supp \text{ } f $, $\supp \text{ } g \subseteq B_r$, where $r = N\epsilon$,  \\\indent (ii) $|\hat{f}|^2 = |\hat{g}|^2 \nequiv 0 \text{ on } \mathbb{R}^d$, \\
\indent (iii) $f \neq g$ in the sense of distributions on $\mathbb{R}^d$, moreover, $f$ and $g$ are not associated functions in the sense of formulas (2), (3),\\
\indent (iv) in addition, if $N \geq d$ and there are $d$ linearly independent $\omega_j$ in (25), (26), then $f$, $g \in L^2(B_r)$.}
\\ \hfill \\
\indent Theorem 3.1 is of interest for $d \geq 2$, whereas it reduces to known results for $d = 1$. \\
\indent Results like Theorem 3.1 are also known for the case when $N = d$ and $\omega_j \cdot \omega_k = 0$ for $j \neq k$, and are sometimes referred as a construction of non-uniqueness for phase retrieval in multidimensions via "tensorization". The point is that Theorem 3.1 strongly generalizes this known "tensorization" construction via "oblique tensorization" arising in (25), (26). \\
\indent Theorem 3.1 is proved in Section 4, using, in particular, some complex analysis in $\mathbb{C}^d$.  
\\ \hfill \\
\indent In turn, Theorem 3.1 can be generalized as follows. 
\\ \hfill \\
\indent \textbf{Theorem 3.2}: \textit{Let $f$, $g$ be defined by
\begin{equation}
     f = F^{-1}\hat{f},\indent  \hat{f}(p) = \prod_{j = 1} ^N \hat{f}_j( p), \text{ } p \in \mathbb{R}^d, 
\end{equation}
\begin{equation}
       g = F^{-1}\hat{g},\indent  \hat{g}(p) = \prod_{j = 1} ^N \hat{g}_j(p),\text{ } p \in \mathbb{R}^d, 
\end{equation}
where $f_j$, $g_j$ are distributions on $\mathbb{R}^d$, $ \supp f_j$, $\supp g_j \subseteq B_{\epsilon_j}$, $\epsilon_j > 0$, and  $|\hat{f}_j|^2 = |\hat{g}_j|^2 \nequiv 0 \text{ on } \mathbb{R}^d$. Then:\\
\indent (i) $\supp \text{ } f $, $\supp \text{ } g \subseteq B_r$, where $r = \sum_{j = 1}^N \epsilon_j$,  \\
\indent (ii) $|\hat{f}|^2 = |\hat{g}|^2 \nequiv 0 \text{ on } \mathbb{R}^d$, \\
\indent (iii) $f$ and $g$ are not associated functions in the sense of formulas (2), (3), under the assumption that the set of zeros (with multiplicity) of $\hat{g}(p)$ in $\mathbb{C}^d$ is different from such for $\hat{f}(p)$ and from such for $\overline{\hat{f}\left(\bar{p}\right)}$.}
\\ \hfill \\
\indent Items (i) and (ii) of Theorem 3.2 are proved in the same way as items (i) and (ii) of Theorem 3.1. \\
\indent Item (iii) of Theorem 3.2 follows from formulas (5), (6). \\
\indent In addition, obviously, $f$, $g \in L^2(B_r)$ if $\hat{f}$, $\hat{g} \in L^2(\mathbb{R}^d)$.\\
\indent Note that the assumption about zeros in item (iii) of Theorem 3.2 is not convenient. Theorem 3.2 is mainly of interest in its particular cases, when these assumptions are fulfilled automatically. \\
\indent One can see that Theorem 3.2 reduces to Theorem 3.1 if the factors $\hat{f}_j$, $\hat{g}_j$ in (27), (28) are of the same form as in (23) - (26). \\
\indent On the other hand, using ideas going back to [13], we can take $\hat{f}$, $\hat{g}$ in (27), (28) as  
\begin{equation}
    \hat{f}(p) =  \hat{a}(p)\hat{b}(p), \indent \hat{g}(p) =  \overline{\hat{a}\left(\overline{p}\right)}\hat{b}(p), \, p \in \mathbb{C}^d,
\end{equation}
 where $\hat{a}$ is a polynomial of $p$, and $a$ is the related distribution, which can be interpreted as a linear differential operator $\mathcal{A}$, whereas $b$ is sufficiently smooth so that $\hat{f}$, $\hat{g} \in L^2(\mathbb{R}^d)$.  
\\ \hfill \\
\indent \textbf{Remark 3.1}: \textit{The functions $\hat{f}$, $\hat{g}$ in (25), (26) can be presented as (27), (28), where 
\begin{equation}
    \hat{a}(p) = \prod_{j = 1}^N \hat{a}_j(\omega_j \cdot p), \indent \hat{a}_j(z)  = \prod_{\zeta \in Z_j} \left( 1 - z/\zeta \right), \, p\in \mathbb{C}^d, z \in \mathbb{C},
\end{equation}
\begin{equation}
    \hat{b}(p) = \frac{\hat{f}(p)}{\hat{a}(p)}=\prod_{j = 1}^N \hat{b}_j(\omega_j \cdot p), \indent \hat{b}_j(z) = \hat{f}_j(z) / \hat{a}_j(z), \, p\in \mathbb{C}^d, z \in \mathbb{C}.
\end{equation}
\indent In addition, if all ${Z}_j$ (mentioned in connection with (23)) are finite, then $\hat{a}$ is a polynomial of $p$ of finite degree. In this sense, Theorem 3.1 can be seen as a particular case of Theorem 3.2, when $\hat{f}$, $\hat{g}$ are presented as in (29). \\
\indent In particular, under the assumptions of item (iv) of Theorem 3.1, $\hat{b}$ in (31) corresponds to a sufficiently smooth $b$ in the sense mentioned after (29). }
\\ \hfill \\
\indent It is important to mention that in contrast to Theorem 3.2, in Theorem 3.1 there is no assumption that the set of zeros (with multiplicity) of $\hat{g}(p)$ in $\mathbb{C}^d$ is different from such for $\hat{f}(p)$ and from such for $\overline{\hat{f}\left(\bar{p}\right)}$. The point is that the proof of Theorem~3.1 includes the proof of this property. In this sense, Theorem 3.1 is a stronger result than Theorem~3.2. 
\\ \hfill \\
\indent In addition to Theorems 3.1 and 3.2, the following result also holds. 
\\ \hfill \\
\indent \textbf{Theorem 3.3}: \textit{Let
\begin{equation}
    f = \mathcal{A}b, \indent g = \mathcal{A^*}b,
\end{equation}
where $\mathcal{A}$ is a linear differential operator with constant coefficients, $\mathcal{A}^*$ is adjoint to $\mathcal{A}$, and $b$ is a smooth compactly supported function on $\mathbb{R}^d$. Suppose that
\begin{equation}
    \mathcal{A^*} \neq e^{i\alpha}\mathcal{A},\, \alpha \in \mathbb{R},
\end{equation}
and $b$ is not associated to itself in the sense of formula (3). Then: \\
\indent (i) $f$ and $g$ are compactly supported functions on $\mathbb{R}^d$,\\
\indent (ii) $|\hat{f}|^2 = |\hat{g}|^2 \nequiv 0 \text{ on } \mathbb{R}^d$, \\
\indent (iii) $f$ and $g$ are not associated functions in the sense of formulas (2), (3). }
\\ \hfill \\
\indent It is important to mention that in Theorem 3.3, we do not have conditions for zeros, as in Theorem 3.2. To satisfy condition (33), one can take, for example, $\mathcal{A} = -\Delta + c\partial_{x_1}$, where $\Delta$ is the Laplacian operator, $c \in \mathbb{R}$.\\
\indent Theorem 3.3 is proved in Section 6. 
\subsection{The case of strongly disconnected support}
\indent In this subsection, we present non-trivial examples of non-uniqueness in phase retrieval for functions with strongly disconnected compact support in multidimensions. 
\\ \hfill \\
\indent \textbf{Remark 3.2}: \textit{Non-trivial examples of non-uniqueness in phase retrieval for functions with strongly disconnected compact support in multidimensions can be constructed, proceeding from Theorem 3.2 for $N = 2$,  where
\begin{equation}
    f_2(x) = \sum_{k = 1}^n C_k \delta(x - y_k),\indent g_2(x) = \sum_{k = 1}^n C_k' \delta(x - y_k), 
\end{equation}
 and $C_k$, $C_k' \in \mathbb{C}$, $y_k \in \mathbb{Z}^d$, $\supp \, f_1$, $\supp \, g_1 \subseteq B_\epsilon$, $\epsilon < 1/2$. \\
\indent For the case when $d = 1$ and $f_1 = g_1$, this construction is presented in detail in Example 1 in [7]. }
\\ \hfill \\
\indent We consider complex-valued functions $v$ on $\mathbb{R}^d$ of the form  
\begin{equation}
    v = \sum_{k = 1}^n v_k \text{, } \supp \text{ }v_k \subset D_k \text{, } \hat{v}_k \nequiv 0,
\end{equation}
 where 
\begin{equation}
    D_k \text{ are open convex bounded domains in } \mathbb{R}^d \text{, } \mathrm{dist}(D_i, D_j) \geq r > 0 \text{ for }i \neq j.
\end{equation}
\indent Here, $\mathrm{dist}(\mathcal{A}, \mathcal{B})$ denotes the distance between sets $\mathcal{A}$ and $\mathcal{B}$ in $\mathbb{R}^d$. \\
\indent Let
\begin{equation}
    N_{\epsilon}(U) = \{ x \in \mathbb{R}^d: \mathrm{dist}(x, U) < \epsilon\},\text{ } \epsilon > 0 \text{, }U \subset \mathbb{R}^d.
\end{equation}
\\ \hfill \\
\indent \textbf{Theorem 3.4}: \textit{Let $v$ be as in (35), (36). Let $f$, $g$ be two complex-valued functions on $\mathbb{R}^d$ such that $\supp \text{ } f$, $\supp \text{ } g \subseteq  B_\delta$, $\delta < r/2$,  and $|\hat{f}|^2 = |\hat{g}|^2 \nequiv ~0 \text{ on } \mathbb{R}^d$, but $f$ and $g$ are not associated functions in the sense of formulas (2), (3). Let $v_f = f * v$, $v_g = g * v$. Then: \\
\indent (i) $v_f$, $v_g$ are of the form (35), (36) with $N_\delta(D_k)$ in place of $D_k$ and $r_\delta = r - 2 \delta$ in place of $r$, \\
\indent (ii)  $|\hat{v}_f|^2 = |\hat{v}_g|^2 \nequiv 0$ on $\mathbb{R}^d$, \\
\indent (iii) $v_f \neq v_g$, moreover, $v_f$ and $v_g$ are not associated functions in the sense of formula (2), \\
\indent (iv) $v_f$ and $v_g$ are not associated functions in the sense of formula (3), under the additional condition that the set of zeros (with multiplicity) of $\hat{g}(p)\hat{v}(p)$ in $\mathbb{C}^d$ is different from such for $\overline{\hat{f}(\bar{p})\hat{v}(\bar{p})}$. \\
\indent (v) $v_f$ and $v_g$ are not associated functions in the sense of formula (3) if, for example, $v(x) = \overline{v(-x)}$, $x \in \mathbb{R}^d$. }
\\ \hfill \\
\indent Theorem 3.4 is proved in Section 5. \\
\indent Note that Theorem 3.1 gives a large class of possible functions $f$, $g$ for Theorem 3.4. \\
\indent In particular, one can consider Theorem 3.4 assuming that $f$, $g \in L^2(B_{\delta})$, $v_k \in L^2(D_k)$.  \\
\indent In addition, Theorem 3.4 is of interest even when all $v_k$ in (35) are Dirac delta functions, i.e., 
\begin{equation}
    v_k(x) = C_k \delta(x - y_k), \text{ }y_k \in D_k, \text{ }C_k \in \mathbb{C}, \text{ }k = 1, \cdots, n. 
\end{equation}
\indent Theorem 3.4, for $n \geq 2$, gives an interesting class of non-trivial examples of non-uniqueness in phase retrieval for functions with strongly disconnected compact support in multidimensions, taking also into account Theorem 3.1.  \\
\indent Theorem 3.4 is of some interest even when $n = 1$ in (35). \\
\indent The non-uniqueness in phase retrieval for functions with strongly disconnected support given by Theorems 3.1 and 3.4 is already of interest when $N = 1$ in (25), (26) and $v_k \in L^2(D_k)$. In this case, $f$, $g$, $v_f$, $v_g$ can be written as follows:
\begin{equation}
    f(x) = (2\pi)^{d - 1}f_1(\omega_1x) \delta(x - (\omega_1x)\omega_1), \indent 
    g(x) = (2\pi)^{d - 1} g_1(\omega_1x) \delta(x - (\omega_1x)\omega_1), \text{ } x \in \mathbb{R}^d,
\end{equation}
\begin{equation}
    v_f(x)= (2\pi)^{d-1} \int_{\mathbb{R}} f_1(\omega_1 x - s) v(x + (s - \omega_1 x)\omega_1  )  ds, \text{ } x \in \mathbb{R}^d,
    \end{equation}
    \begin{equation}
    v_g(x) = (2\pi)^{d-1} \int_{\mathbb{R}} g_1(\omega_1 x - s) v(x + (s - \omega_1 x)\omega_1  )   ds , \text{ } x \in \mathbb{R}^d,
    \end{equation}
 where $\omega_1 \in \mathbb{S}^{d-1}$, $f_1$, $g_1$ are as in formula (23), (24) for $j = 1$, $v$ is as in formulas (35), (36). \\
 \indent The non-uniqueness in phase retrieval given by formulas (39) - (41) is illustrated numerically in [12]. 
\\ \hfill \\
\indent Let 
\begin{equation}
    D = \bigcup_{k = 1}^n D_k,
\end{equation}
where $D_k$ are arbitrary disconnected non-empty bounded domains in $\mathbb{R}^d$, $\mathrm{dist}(D_i, D_j) \geq r > 0$ for $i \neq j$. 
\\ \hfill \\
\indent \textbf{Theorem 3.5}: \textit{Let $f$, $g$ and $\mathcal{A}$, $b$ be as in (32), where $\mathcal{A}$ satisfies (33). Suppose that $\supp \, b = \overline{D}$, where $D$ is defined as in (42), $\overline{D}$ is the closure of $D$, and\\
\begin{equation}
    \overline{D} \neq -\overline{D} + y, \, y \in \mathbb{R}^d.
\end{equation}
\indent Then: \\
\indent (i) $b$ is not associated to itself in the sense of formula (3), \\
\indent (ii) $f$ and $g$ have the properties (i) - (iii) of Theorem 3.3, \\
\indent (iii) $\supp \, f$, $\supp \, g \subseteq \overline{D}$, $\supp \, f \, \cap \overline{D_k} \neq \emptyset$, $\supp \, g \, \cap \overline{D_k} \neq \emptyset$ for each $k$, i.e., $f$, $g$ have strongly disconnected supports for $n \geq 2$. }
\\ \hfill \\
\indent To satisfy condition (43), one can take, for example, $\overline{D} = B_{r_1}(a_1) \cup B_{r_2}(a_2)$, where $B_r(a)$ is defined as in (22), $|a_1 - a_2| > r_1 + r_2$, and $r_1 \neq r_2$.\\
\indent Theorem 3.5 is proved in Section 7.
\section{Proof of Theorem 3.1}
\indent Item (i) follows from formulas (7), (19), (21) with $a_1 = a_2 = 0$. One can also obtain this result proceeding from assumptions (23), definitions (25), (26) and Theorem 2.1.\\
\indent Item (ii) follows from definitions (25), (26) and the property that $|\hat{f}_{j}|^2 = |\hat{g}_{j}|^2 \nequiv 0 $ on $\mathbb{R}$, $j = 1, \cdots, N$.\\
\indent The proof of item (iv) is as follows. \\
\indent Without restriction of generality, we can assume that $\omega_1, \cdots \omega_d$ are linearly independent. \\
\indent Let
\begin{equation}
    \phi(p) = \prod_{j=1}^d \hat{f}_j(\omega_j \cdot p), \indent \psi(p) = \prod_{j=1}^d \hat{g}_j(\omega_j \cdot p), \text{ } p \in \mathbb{R}^d.
\end{equation}
\indent Then using that $f_j$, $g_j \in L^2(\mathbb{R})$ and using the change of variables $\tilde{p}_j = \omega_j \cdot p$, $j = 1, \cdots, d$, we get 
\begin{equation}
    \phi, \psi \in L^2(\mathbb{R}^d). 
\end{equation}
\indent We also have that 
\begin{equation}
    \hat{f}_j(\omega_j \cdot p), \text{ }\hat{g}_j(\omega_j \cdot p) \text{ are analytic and bounded on } \mathbb{R}^d \text{, } j = 1, \cdots, N,
\end{equation}
since $f_j$, $g_j \in L^2(\mathbb{R})$ and $\supp \text{ } f_{j} $, $\supp \text{ } g_{j} \subseteq [-\epsilon, \epsilon]$. \\
\indent Definitions of $\hat{f}$, $\hat{g}$ in (25), (26) and formulas (44) - (46) imply that $\hat{f}, \hat{g} \in L^2(\mathbb{R}^d)$. Thus, item (iv) is proved.\\
\indent The proof of item (iii) is as follows. \\
\indent Since $f$ and $g$ are compactly supported, to prove that $g$ is not associated to $f$ in the sense of (5) and (6), respectively, it is sufficient to show that 
\begin{equation}
    Z_u \cup Z_{{1}/{u}} \neq \emptyset, \indent u(p) = \hat{g}(p)/\hat{f}(p), \, p \in \mathbb{C}^d,
\end{equation}
\begin{equation}
     Z_u \cup Z_{{1}/{u}} \neq \emptyset, \indent u(p) = \hat{g}(p)/\overline{\hat{f}\left(\overline{p}\right)}, \, p \in \mathbb{C}^d,
\end{equation}
respectively, where $Z_u$ denotes the set of zeros of a meromorphic function $u$ in $\mathbb{C}^d$. 
\\
\indent In the proofs below, we will use that 
\begin{equation}
    Z_h \cup Z_{{1}/{h}} \neq \emptyset, \indent h(z) = \hat{g}_j(z)/\hat{f}_j(z), \, z \in \mathbb{C},
\end{equation}
\begin{equation}
   Z_h \cup Z_{{1}/{h}} \neq \emptyset, \indent h(z) = \hat{g}_j(z)/\overline{\hat{f}_j\left(\overline{z}\right)}, \, z \in \mathbb{C},
\end{equation}
 where $Z_h$ denotes the zeros of a meromorphic function $h$ in $\mathbb{C}$ and $j = 1, \cdots N$. This follows from (24). \\
\indent Let us use the presentations, for a fixed $j$:
\begin{equation}
    \hat{f}(p) = {\hat{f}_j(\omega_j \cdot p)}\phi_j(p), \indent  \hat{g}(p) = {\hat{g}_j(\omega_j \cdot p)}\psi_j(p), 
\end{equation}
\begin{equation}
    \phi_j(p) = \prod_{i=1, \, i \neq j}^N \hat{f}_i(\omega_i \cdot p), \indent \psi_j(p) = \prod_{i=1, \,i \neq j}^N \hat{g}_i(\omega_i \cdot p), \text{ } p \in \mathbb{C}^d.
\end{equation}
\indent Let 
\begin{equation}
    Z_{\omega, \zeta} = \{ p \in \mathbb{C}^d\text{ }: \text{ } \omega \cdot p = \zeta\}, \text{ }\omega \in \mathbb{S}^{d-1}, \, \zeta \in \mathbb{C}.
\end{equation}
\indent Note that 
\begin{equation}
    Z_{\omega, \overline{\zeta}} = \overline{Z_{\omega, \zeta}},
\end{equation}
\begin{equation}
   \mathrm{dim}_{\mathbb{C}}\left( Z_{\omega, \zeta} \cap Z_{\theta, \eta} \right) = d - 2, \, \theta \neq \pm \omega,
\end{equation}
\begin{equation}
    Z_{\omega, \zeta} \cap Z_{\omega, \eta} = \emptyset, \, \eta \neq \zeta,
\end{equation}
 where $\omega$, $\theta \in \mathbb{S}^{d-1}$ and $\zeta$, $\eta \in \mathbb{C}$. \\
\indent The proof of (47) is as follows. We take $\zeta \in Z_h \cup Z_{{1}/{h}}$ defined in (49). We consider the case $\zeta \in  Z_h \neq \emptyset$ (the case $\zeta \in  Z_{1/h} \neq \emptyset$ is similar).\\
\indent We have that 
\begin{equation}
    \frac{\hat{g}(p)}{\hat{f}(p)} = \frac{{\hat{g}_j(\omega_j \cdot p)}}{\hat{f}_j(\omega_j \cdot p)}\frac{\psi_j(p)}{\phi_j(p)}, \, p \in \mathbb{C}^d.
\end{equation}
\indent Using the definition of $Z_h$ in (49), formulas (52), (53), (55), and the property that $Z_{\hat{f}_i}$ are countable, one can see that 
\begin{equation}
    Z_{\omega_j, \zeta} \subseteq Z_u, \indent u(p) = \hat{g}_j(\omega_j \cdot p)/\hat{f}_j(\omega_j \cdot p), \, p \in \mathbb{C}^d,
\end{equation}
\begin{equation}
    Z_{\omega_j, \zeta} \nsubseteq Z_{\phi_j} = \bigcup_{i = 1, i \neq j }^N\left( \bigcup_{\zeta \in Z_{\hat{f}_i}} Z_{\omega_i, \zeta}\right),
\end{equation}
 where $Z_{\hat{f}_i}$ denotes the set of zeros of $\hat{f}_i$ in $\mathbb{C}$. \\
\indent Using (57) - (59), we get that $Z_u \neq \emptyset$, where $u$ is as in (47). This completes the proof of (47) for $\zeta \in Z_h \neq \emptyset$. \\
\indent The proof of (48) is similar to that of (47), using also (50) and (54). \\
\indent This completes the proof of item (iii). Theorem 3.1 is proved. 
\section{Proof of Theorem 3.4}
\indent Note that 
\begin{equation}
    v_f = f* v = \sum_{k = 1}^n f * v_k, \indent v_g = g*v =\sum_{k = 1}^n g * v_k.
\end{equation}
\indent Due to formula (8), we have also that
\begin{equation}
    \hat{v}_f = (2\pi)^{-d}\hat{f} \hat{v}, \indent \hat{v}_g = (2\pi)^{-d}\hat{g} \hat{v}.
\end{equation}
\indent Recall also that
\begin{equation}
    \operatorname{mes} \left( \{ p \in \mathbb{R}^d \text{ }: \text{ } \hat{u}(p) = 0\}\right) = 0,
\end{equation}
 where $\hat{u}$ is a non-zero real analytic function on $\mathbb{R}^d$, for example, the Fourier transform of a non-zero compactly supported function $u$ on $\mathbb{R}^d$, and $\operatorname{mes}$ denotes the Lebesgue measure in $\mathbb{R}^d$. \\
\indent In view of (5), (6), (61), we have also that $v_f$ and $v_g$ are associated functions in the sense of formula (2) if and only if 
\begin{equation}
    \hat{g}(p)\hat{v}(p) = \hat{f}(p) \hat{v}(p)\operatorname{exp}(i\alpha + iyp) \text{, } p \in \mathbb{R}^d,
\end{equation}
and in the sense of formula (3) if and only if
\begin{equation}
    \hat{g}(p)\hat{v}(p) = \overline{\hat{f}(p)\hat{v}(p)}\operatorname{exp}(i\alpha + iyp) \text{, } p \in \mathbb{R}^d.
\end{equation}
\indent Item (ii) follows from (61), (62) and the assumptions that $f$, $g$, $v$ are compactly supported, $|\hat{f}|^2 = |\hat{g}|^2 \nequiv 0\text{ on } \mathbb{R}^d$ and $\hat{v} \nequiv 0$ on $\mathbb{R}^d$. \\
\indent Item (iii) follows from (62), (63) and the assumptions that $f$, $g$, $v$ are compactly supported, $\hat{f}$, $ \hat{g}$ are not associated functions in the sense of formulas (2) and $\hat{v} \nequiv 0$ on $\mathbb{R}^d$. \\
\indent Item (iv) follows from the holomorphic extension of (64) into $\mathbb{C}^d$.  \\
\indent Note that 
\begin{equation}
    \hat{v}(p) = \overline{\hat{v}(p)}, \, p \in \mathbb{R}^d \text{ if } v(x) = \overline{v(-x)}, \, x \in \mathbb{R}^d.
\end{equation}
\indent Item (v) follows from (62), (64), (65) and the assumptions that $f$, $g$, $v$ are compactly supported, $\hat{f}$, $ \hat{g}$ are not associated functions in the sense of formulas (3) and $\hat{v} \nequiv 0$ on $\mathbb{R}^d$. \\
\indent The properties that $\supp \text{ } f$, $\supp \text{ } g \subseteq  B_\delta$, $\supp \text{ }v_k \subset D_k$ and formula (19) imply that 
\begin{equation}
    \supp \text{ } f * v_k \subset N_\delta(D_k), \indent  \supp \text{ } g * v_k \subset N_\delta(D_k). 
\end{equation}
\indent In addition, since $\mathrm{dist}(D_i, D_j) \geq r > 2\delta, \text{ } i \neq j$, we have that 
\begin{equation}
    \mathrm{dist}(N_{\delta}(D_i), N_{\delta}(D_j)) \geq r_\delta, \text{ } i \neq j.
\end{equation}
\indent Item (i) follows from formulas (60), (66) and (67). \\
\indent This completes the proof of Theorem 3.4. 
\section{Proof of Theorem 3.3}
\indent Item (i) follows from the definitions in (32). \\
\indent Item (ii) follows from (29), (62) and the observation that, under our assumptions,  
\begin{equation}
    \hat{a}, \hat{b} \text{ are non-zero analytic functions}
\end{equation}
(i.e., more precisely, real analytic on $\mathbb{R}^d$ and holomorphic on $\mathbb{C}^d$).\\
\indent The proof of item (iii) is as follows. \\
\indent In view of (5), (6), we have that $f$ and $g$ are associated functions in the sense of formula (2) if and only if 
\begin{equation}
    \overline{\hat{a}\left(\overline{p}\right)}\hat{b}(p) = \hat{a}(p) \hat{b}(p)\operatorname{exp}(i\alpha + iyp) \text{, } p \in \mathbb{C}^d,
\end{equation}
and in the sense of formula (3) if and only if
\begin{equation}
    \overline{\hat{a}\left(\overline{p}\right)}\hat{b}(p) = \overline{\hat{a}\left(\overline{p}\right) \hat{b}\left(\overline{p}\right)}\operatorname{exp}(i\alpha + iyp) \text{, } p \in \mathbb{C}^d,
\end{equation}
for some $\alpha \in \mathbb{R}$ and $y \in \mathbb{R}^d$. \\
\indent Note that assumption (33) can be rewritten as 
\begin{equation}
\overline{a\left(\overline{p}\right)} \neq e^{i\alpha} a(p), \, p \in \mathbb{C}^d,
\end{equation}
whereas the assumptions that $b$ is compactly supported and is not associated to itself in the sense of formula (3) implies that 
\begin{equation}
    \hat{b}(p) \neq \overline{\hat{b}\left(\overline{p}\right)} \operatorname{exp}(i\alpha + iyp), \, p \in \mathbb{C}^d,
\end{equation}
for any $\alpha \in \mathbb{R}$ and $y \in \mathbb{R}^d$. \\
\indent One can see that equality (69) is impossible, using (62), (68), (71), and using that $\hat{a}$ is a polynomial. \\
\indent One can see that equality (70) is impossible, using (62), (68), (72). \\
\indent Theorem 3.3 is proved. 
\section{Proof of Theorem 3.5}
\indent Item (i) follows from assumption (43) on $\supp \, b = \overline{D}$ and the observation that if $b$ is associated to itself in the sense of formula (3), then 
\begin{equation}
    \supp \, b = - \supp \, b + y \text{ for some } y \in \mathbb{R}^d.
\end{equation} 
\indent Item (ii) follows from item (i) and Theorem 3.3. \\
\indent The proof of item (iii) is as follows. \\
\indent The properties that $\supp \, f$, $\supp \, g \subseteq \overline{D}$ follow from (32) and the assumption that $\supp \, b = \overline{D}$. \\
\indent To prove that $\supp \, f \, \cap \overline{D_k} \neq \emptyset$, $\supp \, g \, \cap \overline{D_k} \neq \emptyset$, it is sufficient to show that 
\begin{equation}
    \mathcal{A}b_k \nequiv 0, \indent \mathcal{A^*}b_k \nequiv 0,
\end{equation}
where $b_k =  \chi_kb$, $\chi_k$ is the characteristic function of $\overline{D_k}$. \\
\indent In the Fourier domain, (74) can be rewritten as 
\begin{equation}
    \hat{a}(p)\hat{b}_k (p)\nequiv 0, \indent \overline{\hat{a}\left(\overline{p}\right)}\hat{b}_k (p)\nequiv 0, \, p\in \mathbb{C}^d.
\end{equation}
\indent Properties (75) follow from the properties that $\hat{a}$, $\hat{b}_k$ are analytic and are not identically zero. \\
\indent In addition, to see that $\hat{b}_k \nequiv 0$, one can use that $\supp \, b_k = \overline{D_k}$ in view of our assumption on $\supp \, b$.\\
\indent Theorem 3.5 is proved. 
\section*{Acknowledgments}
\indent We thank Professor Ph. Jaming for remarks and references that helped to improve strongly this article. In particular, we thank him for indicating us Example 1 in [7] and its possible extension to multidimensions.
\section*{References}
[1] R. H. T. Bates, M. J. McDonnell,  "Image restoration and reconstruction", Oxford University Press (1986) 

[2] T. R. Crimmins, J. R. Fienup, "Ambiguity of phase retieval for functions with disconnected supports", Journal of the Optic Society of America, 71, 1026-1028 (1981)

[3] T. R. Crimmins, J. R. Fienup, "Uniqueness of phase retrieval for functions with sufficiently disconnected support~", Journal of the Optic Society of America, 73, 218 - 221 (1983)

[4] E. M. Hofstetter, "Construction of time-limited functions with specified autocorrelation functions," IEEE Trans. Inf. Theory IT-10, 119-126 (1964)

[5] L. Hörmander, Linear Partial Differential Operators, Volume 1, Springer (1976) 

[6] T. Hohage, R. G. Novikov, V. N. Sivkin, "Phase retrieval and phaseless inverse scattering with background information", Inverse Problems, 40(10), 105007 (2024)

[7] Ph. Jaming, "Phase retrieval techniques for radar ambiguity problems", Journal of Fourier Analysis and Applications, 5, 313-333 (1999)

[8] M. V. Klibanov, P. E. Sacks, A. V. Tikhonravov, "The phase retrieval problem", Inverse Problems, 11(1), 1-28 (1995)

[9] B. Leshem, R. Xu et al., "Direct single shot phase retrieval from the diffraction pattern of separated objects", Nature Communications 7(1), 1-6 (2016) 

[10] J. L. Lions, "Supports dans la transformation de Laplace", J. Analyse Math., 2, 369-380 (1952)  

[11] R. G. Novikov, V. N. Sivkin, "Phaseless inverse scattering with background information", Inverse Problems,
37(5), 055011 (2021)

[12] R. G. Novikov, T. Xu, "On non-uniqueness of phase retrieval in multidimensions", hal-04692589v2  

 [13] J. Rosenblatt, "Phase retrieval", Communications in mathematical physics, 95, 317-343 (1984)  

[14] E. Titchmarsh, "The Theory of Functions", 2nd ed., Oxford University Press, London (1939)

[15] A. Walther, "The question of phase retrieval in optics", International Journal of Optics, 10:1, 41-49 (1963)
\\ \hfill \\
Roman G. Novikov, CMAP, CNRS, Ecole polytechnique, \\
Institut Polytechnique de Paris, 91120 Palaiseau, France\\
\& IEPT RAS, 117997 Moscow, Russia \\
E-mail: novikov@cmap.polytechnique.fr
\\ \hfill \\
Tianli Xu, Ecole polytechnique, \\
Institut Polytechnique de Paris, 91120 Palaiseau, France\\
E-mail: tianli.xu@polytechnique.edu 
 
\end{document}